\documentstyle[preprint,aps,floats,amssymb,epsfig]{revtex}

\newcommand{\be}{\begin{equation}}
\newcommand{\ee}{\end{equation}}
\newcommand{\bea}{\begin{eqnarray}}
\newcommand{\eea}{\end{eqnarray}}
\newcommand{\bml}{\begin{mathletters}}
\newcommand{\eml}{\end{mathletters}}

\newcounter{fixy}
\begin{document}
\newenvironment{fixy}[1]{\setcounter{figure}{#1}}
{\addtocounter{fixy}{1}}
\renewcommand{\thefixy}{\arabic{fixy}}
\renewcommand{\thefigure}{\thefixy\alph{figure}}
\setcounter{fixy}{1}
\tighten

\preprint{DCPT-03/17}
\draft




\title{Electrons on Hexagonal lattices and applications to nanotubes}
\renewcommand{\thefootnote}{\fnsymbol{footnote}}

\author{Betti Hartmann\footnote{Betti.Hartmann@durham.ac.uk}}
\address{Department of Mathematical Sciences, University
of Durham, Durham DH1 3LE, U.K.}
\author{Wojtek J. Zakrzewski\footnote{W.J.Zakrzewski@durham.ac.uk}}
\address{Department of Mathematical Sciences, University
of Durham, Durham DH1 3LE, U.K.}
\date{\today}
\setlength{\footnotesep}{0.5\footnotesep}

\maketitle
\begin{abstract}
We consider a Fr\"ohlich-type Hamiltonian on a hexagonal lattice.
Aiming to describe nanotubes, we choose this
 2-dimensional lattice to be periodic and to have a large extension
in one ($x$) direction and a small extension in the other ($y$) direction. 
We study the existence of solitons in this model using both analytical
and numerical methods. We find exact solutions of our equations and discuss 
some of their properties.
\end{abstract}
\pacs{PACS numbers: 05.45.Yv, 61.46.+w, 63.20.Kr, 81.07.De }
\renewcommand{\thefootnote}{\arabic{footnote}}
\section{Introduction}
Nanotubes have attracted a large amount of interest ever since they were
first discovered in 1991 \cite{iijima}. They can be thought
of as carbon cylinders with a hexagonal grid and are thus 
fullerene related structures.  Their 
mechanical, thermal, optical and electrical properties have been studied 
in some detail \cite{nano}. It was found that most properties depend
crucially on the diameter, chirality and lengths of the tube.
A distortion of the lattice thus affects the energy band gap.
This distortion of the lattice can be achieved in two different
ways: a) through an external force like e.g. bending, stretching or twisting \cite{cn}
or  b) through  an internal excitation, which interacts
with the lattice. It is well known that the interaction of
an excitation such as an amide I- vibration in biopolymers
or an electron (in the case of the Fr\"ohlich Hamiltonian)
with a lattice whose distortion is initially caused by the excitation
results in the creation of a localised state which, in what follows, we refer to as 
a soliton. Such a soliton was first 
introduced by Davydov \cite{davy} in the 1970s  to explain the
dispersion free energy transport in biopolymers (see also \cite{review} for
further details).

Recently, a Fr\"ohlich Hamitonian was studied on 
a two-dimensional, discrete, quadratic lattice \cite{bpz1,bpz3,bpz2}. 
In \cite{bpz1,bpz3}, the existence of localised states 
was studied numerically and it was found that their properties
depend crucially on the electron-phonon coupling constant.
An analytical study confirmed these results \cite{bpz2} - by showing that in the continuum
limit the set of discrete equations reduces to a modified non-linear Schr\"odinger (MNLS) equation
which has an additional term resulting from the discreteness of the lattice.
Although a soliton of the basic NLS 
is unstable the extra term was shown to stabilise it for
appropriate choices of the coupling constant. 

In this paper, we extend the study of Refs. \cite{bpz1,bpz3,bpz2} to the case of
a hexagonal, periodic lattice with a large extension in $x$ and a small extension in
$y$ directions. We study the resultant equations both analytically and numerically.
In Section II we present the Hamiltonian and the equations of motion. In Section III we
discuss various properties of the equations in the stationary limit  and we demonstrate
the existence of an {\bf exact} solution of the discrete equations. In this limit, 
we can thus replace the
full system of equations by a modified discrete non-linear Schr\"odinger (DNLS) equation.
In Section IV, we present and compare our numerical results for the continuous MNLS
equation, for the full system of equations
and for the modified DNLS equation. Most of our numerical simulations
are performed for the $(5,5)$ armchair tube. Such nanotube
was discussed in a recent paper by Liu et al.  in the context of nanorings \cite{liu}.
We also discuss briefly
how nanotubes with different diameter and chirality could be constructed
in our model.

\section{The Hamiltonian and equations of motion}

\subsection{Hamiltonian}
The Hamiltonian $H$ of our model is a sum of four sums which result from
the special features of the hexagonal grid. $\psi_{i,j}$ denotes the electron field on the $i$-th, $j$-th lattice side, while
$u_{i,j}$ and $v_{i,j}$ are the displacements of the $i$-th, $j$-th lattice point from equilibrium in the $x$ 
direction and $y$ directions, respectively:
\begin{eqnarray}
H &=& \sum_{\frac{j-1}{2}=0}^{\frac{N_2}{2}-1} \sum_{\frac{i-1}{4}=0}^{\frac{N_1}{4}-3} \left[ (E+W)\psi_{i,j}\psi^*_{i,j}-
j_x\psi^*_{i,j}\left(\psi_{i+1,j+1}+\psi_{i-1,j}+\psi_{i+1,j-1}\right) \right. \nonumber \\ 
&-& 
\left. j_x\psi_{i,j}\left(\psi^*_{i+1,j+1}+\psi^*_{i-1,j}+\psi^*_{i+1,j-1}\right) \right. \nonumber \\
&+& 
\left. \vert\psi_{i,j}\vert^2\left(\frac{c_x}{3}(u_{i+1,j+1}+u_{i+1,j-1}-2u_{i-1,j})
+\frac{c_x}{\sqrt{3}}(v_{i+1,j+1}-v_{i+1,j-1})\right)\right] \nonumber \\
&+& 
\sum_{\frac{j}{2}=1}^{\frac{N_2}{2}} \sum_{\frac{i-2}{4}=0}^{\frac{N_1}{4}-2} \left[ (E+W)\psi_{i,j}\psi^*_{i,j}-
j_x\psi^*_{i,j}\left(\psi_{i+1,j}+\psi_{i-1,j+1}+\psi_{i-1,j-1}\right) \right. \nonumber \\
&-& 
 \left. j_x\psi_{i,j}\left(\psi^*_{i+1,j}+\psi^*_{i-1,j+1}+\psi^*_{i-1,j-1}\right) \right. \nonumber \\
&+&  
\left. \vert\psi_{i,j}\vert^2\left(\frac{c_x}{3}(-u_{i-1,j-1}-u_{i-1,j+1}+2u_{i+1,j})
+\frac{c_x}{\sqrt{3}}(v_{i-1,j+1}-v_{i-1,j-1})\right)\right] \nonumber  \\
&+& 
\sum_{\frac{j}{2}=1}^{\frac{N_2}{2}} \sum_{\frac{i-3}{4}=0}^{\frac{N_1}{4}-1} \left[(E+W)\psi_{i,j}\psi^*_{i,j}-
j_x\psi^*_{i,j}\left(\psi_{i+1,j+1}+\psi_{i-1,j}+\psi_{i+1,j-1}\right) \right. \nonumber \\
&-& 
\left. j_x\psi_{i,j}\left(\psi^*_{i+1,j+1}+\psi^*_{i-1,j}+\psi^*_{i+1,j-1}\right) \right. \nonumber  \\
&+& 
\left. \vert\psi_{i,j}\vert^2\left(\frac{c_x}{3}(u_{i+1,j+1}+u_{i+1,j-1}-2u_{i-1,j})
+\frac{c_x}{\sqrt{3}}(v_{i+1,j+1}-v_{i+1,j-1})\right)\right] \nonumber \\
&+& 
\sum_{\frac{j-1}{2}=0}^{\frac{N_2}{2}-1} \sum_{\frac{i}{4}=1}^{\frac{N_1}{4}} \left[(E+W)\psi_{i,j}\psi^*_{i,j}-
j_x\psi^*_{i,j}\left(\psi_{i+1,j}+\psi_{i-1,j+1}+\psi_{i-1,j-1}\right)  \right. \nonumber \\
&-& 
\left. j_x\psi_{i,j}\left(\psi^*_{i+1,j}+\psi^*_{i-1,j+1}+\psi^*_{i-1,j-1}\right) \right. \nonumber \\
&+& 
\left. \vert\psi_{i,j}\vert^2\left(\frac{c_x}{3}(-u_{i-1,j-1}-u_{i-1,j+1}+2u_{i+1,j})+
\frac{c_x}{\sqrt{3}}(v_{i-1,j+1}-v_{i-1,j-1})\right)\right]
\end{eqnarray}
with the phonon energy $W$~:
\begin{eqnarray}
W&=&\frac{1}{2}M \sum_{j=1}^{N_2} \sum_{i=1}^{N_1} \left( (\frac{du_{ij}}{dt})^2
+(\frac{dv_{ij}}{dt})^2\right )\nonumber \\
&+&
\frac{1}{2}M \sum_{\frac{j-1}{2}=0}^{\frac{N_2}{2}-1} \sum_{\frac{i-1}{4}=0}^{\frac{N_1}{4}-3} \left(
k_x[(u_{ij}-u_{i-1,j})^2+(v_{ij}-v_{i-1,j})^2 + (u_{ij}-u_{i+1,j+1})^2+(v_{ij}-v_{i+1,j+1})^2 \right. \nonumber \\
&+& 
\left. (u_{ij}-u_{i+1,j-1})^2+(v_{ij}-v_{i+1,j-1})^2] \right) +
\frac{1}{2}M\sum_{\frac{j}{2}=1}^{\frac{N_2}{2}} \sum_{\frac{i-2}{4}=0}^{\frac{N_1}{4}-2} \left(
k_x[(u_{ij}-u_{i+1,j})^2+(v_{ij}-v_{i+1,j})^2 \right.  \nonumber \\
&+&
\left. (u_{ij}-u_{i-1,j+1})^2+(v_{ij}-v_{i-1,j+1})^2
+ (u_{ij}-u_{i-1,j-1})^2+(v_{ij}-v_{i-1,j-1})^2] \right)\nonumber \\
&+&
\frac{1}{2}M \sum_{\frac{j}{2}=1}^{\frac{N_2}{2}} \sum_{\frac{i-3}{4}=0}^{\frac{N_1}{4}-1} \left(
k_x[(u_{ij}-u_{i-1,j})^2+(v_{ij}-v_{i-1,j})^2 + (u_{ij}-u_{i+1,j+1})^2+(v_{ij}-v_{i+1,j+1})^2 \right. \nonumber \\
&+& 
\left. (u_{ij}-u_{i+1,j-1})^2+(v_{ij}-v_{i+1,j-1})^2] \right)+
\frac{1}{2}M\sum_{\frac{j-1}{2}=0}^{\frac{N_2}{2}-1} \sum_{\frac{i}{4}=1}^{\frac{N_1}{4}} \left(
k_x[(u_{ij}-u_{i+1,j})^2+(v_{ij}-v_{i+1,j})^2 \right. \nonumber \\
&+& 
\left.
(u_{ij}-u_{i-1,j+1})^2+(v_{ij}-v_{i-1,j+1})^2
+ (u_{ij}-u_{i-1,j-1})^2+(v_{ij}-v_{i-1,j-1})^2] \right) \ .   
\end{eqnarray}
$j_x$ is the electron field self-interaction coupling, $c_x$ couples
the electron field to the displacement fields $u$ and $v$ and $k_x$ is the
self-coupling of the displacement fields. 
\subsection{Equations of motion}
We can easily derive the equations of motion  from our Hamiltonian $H$.
As an example we give the equations for $i=1+4k$. The discrete Schr\"odinger
equation for the $\psi_{i,j}$ field thus becomes:
\begin{eqnarray}
\label{eq1}
i \hbar\frac{\partial \psi_{i,j}}{\partial t}&=&(E+W)\psi_{i,j}-2j_{x}
\left(\psi_{i+1,j+1}+\psi_{i-1,j}+\psi_{i+1,j-1}\right) \nonumber \\
&+&\psi_{i,j}\left[\frac{c_{x}}{3}(u_{i+1,j+1}+u_{i+1,j-1}-2 u_{i-1,j})
+\frac{c_{x}}{\sqrt{3}}(v_{i+1,j+1}-v_{i+1,j-1})
\right] \ ,
\end{eqnarray}
while the equations for the displacement fields $u_{i,j}$ and $v_{i,j}$ are given by:
\begin{eqnarray}
\label{eq2}
\frac{d^2 u_{i,j}}{dt^2}&=&k_{x}\left(3u_{i,j}-u_{i+1,j+1}-u_{i-1,j}-u_{i+1,j-1}\right)\nonumber \\
&+& \frac{c_{x}}{3M}\left(2\vert\psi_{i-1,j}\vert^2-\vert\psi_{i+1,j+1}\vert^2
-\vert\psi_{i+1,j-1}\vert^2\right) 
\end{eqnarray}
and
\begin{eqnarray}
\label{eq3}
\frac{d^2 v_{i,j}}{dt^2}&=&k_{x}\left(3v_{i,j}-v_{i+1,j+1}-v_{i-1,j}-v_{i+1,j-1}\right)\nonumber \\
&-& \frac{c_{x}}{\sqrt{3}M}\left(\vert\psi_{i+1,j+1}\vert^2 
-\vert\psi_{i+1,j-1}\vert^2\right) \ . 
\end{eqnarray}
We perform  the following rescalings:
\begin{equation}
\tau=\frac{j_x t}{\hbar} \ , \  U=3C_x u \ , \  V=3C_x v \ , \
E_0=\frac{E}{j_x} \ , \ W_0=\frac{W}{j_x}
\end{equation}
and introduce the following rescaled coupling constants:
\begin{equation}
C_x=\frac{c_x}{9 j_x} \ , \  K_x=\frac{k_x \hbar^2}{j_x^2} \ , \ g=\frac{2C_x^2}{E_s} \ , \
E_s=\frac{Mj_x}{9 \hbar^2}  \ .
\end{equation}
The equations then read:
\begin{eqnarray}
i \frac{\partial \psi_{i,j}}{\partial \tau}&=&(E_0+W_0)\psi_{i,j}-2
\left(\psi_{i+1,j+1}+\psi_{i-1,j}+\psi_{i+1,j-1}\right) \nonumber \\
&+&\psi_{i,j}\left[(U_{i+1,j+1}+U_{i+1,j-1}-2 U_{i-1,j})
+\sqrt{3}(V_{i+1,j+1}-V_{i+1,j-1})
\right] \ ,
\end{eqnarray}
\begin{eqnarray}
\frac{d^2 U_{i,j}}{d\tau^2}&=&K_{x}\left(3U_{i,j}-U_{i+1,j+1}-U_{i-1,j}-U_{i+1,j-1}\right)\nonumber \\
&+& \frac{g}{2}\left(2\vert\psi_{i-1,j}\vert^2-\vert\psi_{i+1,j+1}\vert^2
-\vert\psi_{i+1,j-1}\vert^2\right) 
\end{eqnarray}
and
\begin{eqnarray}
\frac{d^2 V_{i,j}}{d\tau^2}&=&K_{x}\left(3V_{i,j}-V_{i+1,j+1}-V_{i-1,j}-V_{i+1,j-1}\right)\nonumber \\
&-& \frac{\sqrt{3}g}{2}\left(\vert\psi_{i+1,j+1}\vert^2 
-\vert\psi_{i+1,j-1}\vert^2\right)    \ .
\end{eqnarray}

\section{Stationary limit}
In the stationary limit, we have:
\begin{eqnarray}
\label{sch}
&&\lambda\psi_{i,j}+2\left(3\psi_{i,j}-\psi_{i+1,j+1}-\psi_{i-1,j}-\psi_{i+1,j-1}\right) \nonumber \\
&+&\psi_{i,j}[U_{i+1,j+1}+U_{i+1,j-1}-2 U_{i-1,j} 
+\sqrt{3}(V_{i+1,j+1}-V_{i+1,j-1})]=0
\end{eqnarray}
with $\lambda=E_0+W_0-6$ and
\begin{eqnarray}
\label{us}
3U_{i,j}-U_{i+1,j+1}-U_{i-1,j}-U_{i+1,j-1}
= -\frac{\tilde{g}}{2}\left(2\vert\psi_{i-1,j}\vert^2-\vert\psi_{i+1,j+1}\vert^2
-\vert\psi_{i+1,j-1}\vert^2\right) 
\end{eqnarray}
\begin{eqnarray}
\label{vs}
3V_{i,j}-V_{i+1,j+1}-V_{i-1,j}-V_{i+1,j-1}
= \frac{\sqrt{3}\tilde{g}}{2}\left(\vert\psi_{i+1,j+1}\vert^2 
-\vert\psi_{i+1,j-1}\vert^2\right) 
\end{eqnarray} 
with $\tilde{g}=\frac{g}{K_x}$.

\subsection{Discrete equation}
In contrast to the square grid, we find that the discrete equations of the hexagonal
grid in the stationary 
limit { do} have an {\it exact} solution. We can thus replace the system of coupled equations (\ref{eq1})-
(\ref{eq3}) by just one modified DNLS equation. 
We again look at  the case $i=1+4k$ for which we have :
\begin{displaymath}
\Delta(1) U_{ij}=\frac{\tilde{g}}{2}\left(2\vert\psi_{i-1,j}\vert^2-\vert\psi_{i+1,j+1}\vert^2-
\vert\psi_{i+1,j-1}\vert^2\right)
\end{displaymath}
where $\Delta(1) U_{ij}=U_{i+1,j+1}+U_{i-1,j}+U_{i+1,j-1}-3U_{i,j}$.
Analogously, we have:
\begin{displaymath}
\Delta(1) V_{ij}=\frac{\sqrt{3}\tilde{g}}{2}\left(\vert\psi_{i+1,j-1}\vert^2-\vert\psi_{i+1,j+1}\vert^2\right)
\end{displaymath}
where $\Delta(1) V_{ij}=V_{i+1,j+1}+V_{i-1,j}+V_{i+1,j-1}-3V_{i,j}$.
Next we note that for the three nearest neighbours we have similar relations, namely:
\begin{displaymath}
\Delta(1) U_{i+1,j+1}=\frac{\tilde{g}}{2}\left(\vert\psi_{i,j}\vert^2+\vert\psi_{i,j+2}\vert^2-
2\vert\psi_{i+2,j+1}\vert^2\right) \  , \
\end{displaymath}
\begin{displaymath}
\Delta(1) U_{i+1,j-1}=\frac{\tilde{g}}{2}\left(\vert\psi_{i,j-2}\vert^2+\vert\psi_{i,j}\vert^2-
2\vert\psi_{i+2,j-1}\vert^2\right) \  , \
\end{displaymath}
\begin{displaymath}
\Delta(1) U_{i-1,j}=\frac{\tilde{g}}{2}\left(\vert\psi_{i-2,j-1}\vert^2+\vert\psi_{i-2,j+1}\vert^2-
2\vert\psi_{i,j}\vert^2\right) \ 
\end{displaymath}
for the $U$ field and
\begin{displaymath}
\Delta(1) V_{i+1,j+1}=
\frac{\sqrt{3}\tilde{g}}{2}\left(\vert\psi_{ij}\vert^2-\vert\psi_{i,j+2}\vert^2\right) \ ,
\end{displaymath} 
\begin{displaymath}
\Delta(1) V_{i+1,j-1}=\frac{\sqrt{3}\tilde{g}}{2}
\left(\vert\psi_{i,j-2}\vert^2-\vert\psi_{ij}\vert^2\right)  
\end{displaymath}
for the $V$ field.

Defining:
$Z_a=U_{i+1,j+1}+U_{i+1,j-1}-2 U_{i-1,j} 
+\sqrt{3}(V_{i+1,j+1}-V_{i+1,j-1})$ (i.e. the lattice terms in (\ref{sch})) we find that the following discrete equation
holds
\begin{eqnarray}
\label{deq}
& &\Delta(1) Z_a =\nonumber \\
& & \tilde{g}\left(6\vert\psi_{ij}\vert^2- \vert\psi_{i,j+2}\vert^2 -\vert\psi_{i+2,j+1}\vert^2-
\vert\psi_{i,j-2}\vert^2 -\vert\psi_{i+2,j-1}\vert^2-\vert\psi_{i-2,j-1}\vert^2 -\vert\psi_{i-2,j+1}\vert^2
\right) \nonumber
\end{eqnarray}
The right hand side of (\ref{deq}) is a 7-point Laplacian
$\Delta(2)\vert\psi_{ij}\vert^2$, thus we find:
\begin{equation}
\label{main}
\Delta(1) Z_a=-\tilde{g}\Delta(2)\vert\psi_{ij}\vert^2 \ .
\end{equation}
It is easy to see that one possible solution of this equation is of the form:
\begin{equation}
\label{solution}
Z_a=-\tilde{g}\left(\vert\psi_{i+1,j+1}\vert^2+\vert\psi_{i+1,j-1}\vert^2+\vert\psi_{i-1,j}
\vert^2+3\vert\psi_{ij}\vert^2\right) \ .
\end{equation}
This is quite remarkable since on a square lattice a similar equation
has no simple solution.
Inserting (\ref{solution})  into (\ref{sch}) we have:
\begin{eqnarray}
\label{NLS}
&&\lambda\psi_{i,j}+2\left(3\psi_{i,j}-\psi_{i+1,j+1}-\psi_{i-1,j}-\psi_{i+1,j-1}\right) \nonumber \\
&-&\tilde{g}\psi_{i,j}[\vert\psi_{i+1,j+1}\vert^2+\vert\psi_{i+1,j-1}\vert^2+\vert\psi_{i-1,j}
\vert^2+3\vert\psi_{ij}\vert^2]=0 \nonumber
\end{eqnarray}
or
\begin{equation}
\label{discrete}
\lambda\psi_{i,j}-2\Delta(1)\psi_{ij}-\tilde{g}\psi_{i,j}\left
(\Delta(1) \vert\psi_{i,j}\vert^2+ 6\vert\psi_{ij}\vert^2\right)=0.
\end{equation}
This equation constitutes our discrete nonlinear Schr\"odinger (DNLS) equation. 

\subsection{Continuum limit}
Next we look at the continuum limit of (\ref{discrete}). To do this we introduce the following expansions:
\begin{eqnarray}
{\psi}_{i\pm 1,j+1}&=&{\psi}\pm \delta x_{\pm}\frac{\partial {\psi}}{\partial x}
+\delta y \frac{\partial {\psi}}{\partial y}+
\frac{1}{2} (\delta x_{\pm})^2 \frac{\partial^2 {\psi}}{\partial x^2}\pm
 \delta y \delta x_{\pm} \frac{\partial^2 {\psi}}{\partial x\partial y}+
\frac{1}{2} (\delta y)^2 \frac{\partial^2 {\psi}}{\partial y^2} \nonumber \\
&\pm&  \frac{1}{8} (\delta x_{\pm})^3 \frac{\partial^3 {\psi}}{\partial x^3}
+  \frac{1}{2} (\delta x_{\pm})^2 \delta y  \frac{\partial^3 {\psi}}{\partial x^2\partial y}
\pm \frac{1}{2} \delta x_{\pm} (\delta y)^2  \frac{\partial^3 {\psi}}{\partial x\partial y^2}
+\frac{1}{8} (\delta y_{\pm})^3 \frac{\partial^3 {\psi}}{\partial y^3} \pm ...
\end{eqnarray}
and
\begin{eqnarray}
{\psi}_{i\pm 1,j-1}&=&{\psi}\pm \delta x_{\pm}\frac{\partial {\psi}}{\partial x}
-\delta y \frac{\partial {\psi}}{\partial y}
+\frac{1}{2} (\delta x_{\pm})^2 \frac{\partial^2 {\psi}}{\partial x^2}\mp
 \delta y \delta x_{\pm} \frac{\partial^2 {\psi}}{\partial x\partial y}+
\frac{1}{2} (\delta y)^2 \frac{\partial^2 {\psi}}{\partial y^2} \nonumber \\
&\pm&  \frac{1}{8} (\delta x_{\pm})^3 \frac{\partial^3 {\psi}}{\partial x^3}
- \frac{1}{2} (\delta x_{\pm})^2 \delta y  \frac{\partial^3 {\psi}}{\partial x^2\partial y}
\pm \frac{1}{2} \delta x_{\pm} (\delta y)^2  \frac{\partial^3 {\psi}}{\partial x\partial y^2}
-\frac{1}{8} (\delta y_{\pm})^3 \frac{\partial^3 {\psi}}{\partial y^3} \pm ...
\end{eqnarray}
where for $i=1+4k$ and $i=3+4k$ we have $\delta x_{+}=1/2$, $\delta x_{-}=1$, while
for $i=2+4k$ and $i=4+4k$ we have $\delta x_{+}=1$, $\delta x_{-}=1/2$. Moreover, $\delta y=\sqrt{3}/2$.
Inserting this into (\ref{discrete}), we obtain:
\begin{equation}
\lambda\psi-\frac{3}{2}\Delta \psi-\tilde{g}\psi\left(\frac{3}{4}\Delta \vert\psi\vert^2+6\vert\psi\vert^2\right)=0
\end{equation}
or, equivalently, 
\begin{equation}
\tilde{\lambda}\psi+\Delta \psi+4\tilde{g}\psi\left(\vert\psi\vert^2+ 
\frac{1}{8}\Delta \vert\psi\vert^2\right)=0
\end{equation}

We thus have, in analogy to what was found in \cite{bpz1,bpz3,bpz2}, a modified nonlinear Schr\"odinger (MNLS) 
equation  with an extra term, which
can stabilise the soliton:
\begin{equation}
\label{cnls}
i\frac{\partial \psi}{\partial \tau}+\Delta \psi +4\tilde{g}\psi\left(\vert\psi\vert^2+ 
\frac{1}{8}\Delta \vert\psi\vert^2\right)=0.
\end{equation}
Following \cite{bpz1,bpz3,bpz2} we see that the conserved energy in this case is given by:
\begin{equation}
{\cal E}=\int \left( |\vec{\nabla}\psi|^2-2\tilde{g} |\psi|^4 + \frac{\tilde{g}}{4}(\vec{\nabla} |\psi|^2)^2\right)dxdy \ .
\end{equation}
Approximating the soliton solution of (21) by a Gaussian of the form $\psi(x,y)=
\frac{\kappa}{\sqrt{\pi}}\exp(-\frac{\kappa^2}{2}(x^2+y^2))$ we find that the value
of $\kappa$ that minimises the energy is
\begin{equation}
\kappa^2_{min}=2(1-\frac{\pi}{\tilde{g}})
\end{equation}
and thus we have an estimate of the critical $\tilde g$, namely,  $\tilde{g}_{cr}\sim\pi$.

\section{Numerical results}
\subsection{Continuous, modified non-linear Schr\"odinger (MNLS) equation }
First, we have considered the continuous, modified non-linear Schr\"odinger equation (\ref{cnls}).
For this, we have taken the radially symmetric  ansatz:
\begin{equation}
\label{ansatz}
\psi(r,t)=e^{i\alpha t} R(r),
\end{equation}
and put this into (21).
Then we have solved the ordinary differential equation (ODE) using a collocation method
for the boundary value ODEs \cite{col} and choosing the boundary conditions:
\begin{equation}
\frac{\partial R(r)}{\partial r}|_{r=0}=0 \ \ , \ \ R(r=\infty)=0 \  .
\end{equation}
Our results are shown in Fig.~1. For a fixed $\tilde{g}$ we have determined the value of the function $R(r)$ at the origin, $R(r=0)$,
as well as the value of $\alpha$ for which the norm of the solution is equal to unity. As can be seen from Fig.~1, we have found
that for a critical value of $\tilde{g}=\tilde{g}_{cr}\approx 2.94$ the value of $R(0)$ tends to zero.
Since our construction is such that the maximum of the solution is located at $r=0$, the value of the height of the solution
tends to zero and thus the solution ceases to exist. 
We thus find
that the critical value of $\tilde{g}$ from our numerical construction  agrees with the
upper-bound value obtained from the variational approach based on the  Gaussian, namely, $\tilde{g}_{cr}=\pi$.

Note that the solutions cease to exist when the value of $\alpha$ tends to zero. 
Since we can interpret $\alpha$ as the frequency of an internal rotation, the solutions apparently cease to exist
when there is no internal rotation. This can be compared to the so-called ``Q-balls'' which are
non-topological solitons characterized  also by a complex scalar field \cite{qballs} of the form similar to
(\ref{ansatz}). For them its is known\cite{qballs} that there exist upper and lower positive-valued bounds on the frequency of
the internal rotation in order for Q-balls to be stable. In comparison, our solutions exist for all values of
$\alpha >0$. This is probably due to the fact that while the dynamical part of
our action is similar to that for Q-balls, we have a extra term involving derivatives as compared to
an ``ordinary'' $\psi^6$-potential in the case of Q-balls.

\subsection{Discrete equations}

\subsubsection{Full system of equations}
For our numerical study of the full equations (\ref{eq1})-(\ref{eq3}) we have found it convenient to ``squeeze'' the lattice
as indicated in Figs.~2a und 2b. The Hamiltonian and the corresponding equations are given in the
Appendix.
For our numerical calculations, we have used mainly a 
periodic grid with $N_1=160$ and $N_2=20$. We have in addition
chosen the boundary conditions such that the fields
at $(i=0,j)$ are identified with those at $(i=i_{max},j)$.
Thus the type of nanotube
we are studying here is a $(5,5)$ armchair tube which is metallic \cite{nano}.
Nanotubes can also be semiconducting and we make a brief comment about 
the possibility of constructing semiconducting tubes in our model
in the last paragraph of this section.

In this work, the C-C bond length, $0.1415$ nm \cite{cbon}, is normalised to unity.
Therefore, the tube diameter is $0.6756$ nm.
Tubes with different diameter
can also be constructed in our model. We discuss this together with different
chiralities in the last paragraph of this section.
 
As starting configuration we have used an exponential-like excitation $\psi_{i,j}$ extended
typically over the lattice points $i=78-83$ and $j=3-7$ with the lattice at equilibrium
everywhere, i.e.  $u_{i,j}=0$ and $v_{i,j}=0$ for all $i,j$.
We are mainly interested 
in the existence of solitons and their dependence on the value of the coupling constant $c_x$. 
We have set $j_x=k_x=1$, $M=20$ and $E=0.142312$. The main goal of this work is to study
the dependence on $c_x$. So the exact values of $j_x$ and $k_x$ play a minor
role. Hence, we have set them to one. The choice of $M=20$
is a reflection of the physical fact that the mass of the carbon atom is approximately $20\cdot 10^{-24} g$.
 
To absorb the energy  thus allowing the initial configuration to evolve into 
the stationary solutions of (\ref{eq1})-(\ref{eq3}), i.e. of (\ref{sch})-(\ref{vs}) we have additionally 
introduced  damping terms $\nu \frac{du_{i,j}}{dt}$ and $\nu \frac{dv_{i,j}}{dt}$, respectively, into
the  equations (\ref{eq2}) and (\ref{eq3}). We have typically chosen $\nu=0.25-0.75$. For this choice of the coupling constants,
we have performed several numerical calculations using a $4$th order Runge-Kutta
method for simulating the time evolution. We have found that solitons exist in this system for $c_x \gtrapprox 20$.
For larger values of $c_x$, the soliton forms very quickly, while decreasing $c_x$
the time increases at which a soliton forms. This is of course due to the
weaker coupling between the dynamics of the lattice itself and the
excitation. For $c_x=19$, we have waited until $t\approx 8000$ and haven't found a soliton.
Moreover, in all cases we have found only little displacement of the lattice from the equilibrium. We 
have found that at the location of the
soliton the lattice becomes squeezed (i.e. the lattice sites move towards
the sites at which the soliton is located). This is demonstrated in Fig.~3 for $c_x=25$, where we show the lattice
distortion after $t=4000$.
The point at which the centre of the soliton is located doesn't move, while 
the sites in its close neighbourhood all move towards the centre of the soliton.

We have also studied the effects of perturbations of the solitons. We have 
found that after perturbing the
soliton we obtain a new solution with a different height of the soliton
maximum.
Even after introducing a perturbation which keeps the maximal height fixed,
the new solution differs from the starting one. We thus come to the conclusion
that the full system of equations (\ref{eq1})-(\ref{eq3}) has a large number of solutions
for each choice of coupling constants. We believe that a conserved quantity exists in this system
which picks out the specific solution. However, so far we have not been able 
 to determine this conserved quantity.

\subsubsection{Modified, discrete non-linear Schr\"odinger (DNLS) equation}
In addition to the full system of equations, we have also studied the dynamical analog of equation (\ref{NLS}).
Using a similar starting configuration with $\psi_{i,j}$ being exponential
and nonzero over $i=78-83$ and $j=3-7$, we have determined the value of $\tilde{g}$ for which a soliton exists.
Our results are shown in Fig.~4 where we present the height of the soliton's maximum $(\psi\psi^*)_{max}$ 
as function of $\tilde{g}$. We find that the value of $\tilde{g}$ at which the soliton
disappears $\tilde{g}_{cr}\approx 2.295$. The height of the soliton at this critical
coupling is $(\psi_{i,j}\psi^*_{i,j})_{max}\approx 0.227$. Our numerical study of the
continuous MNLS equation gave us $\tilde{g}_{cr}\approx 2.94$, while the analytic study led to
$\tilde{g}_{cr}=\pi$. Both values are not a bad approximation
for the value found numerically for the discrete equation.

To test the independence of our results from the form of the initial settings, we
have used a different starting configuration with two exponential-like 
excitations being located at $i=78-83$, $j=3-7$ and $i=138-143$ and $j=13-17$, respectively.
We have found that for values $\tilde{g} \gtrapprox 3$, the results agree. For both types
of initial configurations, the minimal energy configuration corresponds to
 one soliton. However, having said this, 
the time to reach this minimal energy configurations is significantly smaller for the initial
configuration with one excitation than for that with two (typically one order of magnitude smaller). 
We have also tested our results as to the dependence on the size of the grid. For this,
we have chosen two excitations on three different grid sizes: a) a grid with $N_1=160$, $N_2=20$ and
two exponential excitations extended over $i=78-83$, $j=3-7$ and $i=138-143$, $j=13-17$, respectively,
b) a grid with $N_1=320$ and $N_2=40$ with the excitations located at the same places than
in a), and finally c) a grid with $N_1=60$, $N_2=10$ and
two exponential excitations extended over $i=18-23$, $j=2-4$ and $i=38-43$, $j=7-9$, respectively.
We have found that for $\tilde{g}=3$, the results of cases a) and c) agree. 
For the case a) the soliton forms at $t\approx 300$, 
while for the case c) it forms at $t\approx 100$. This is not suprising since in the case c),
the two excitations are located nearer  to each other than in the case a). To test the dependence on
the actual lattice size we have compared the cases a) and b). We have found that the larger the lattice the longer
it takes for the soliton to form. For $\tilde{g}=3$ a soliton forms after $t\approx 300$ in the case a), while
for b) it forms at $t > 700$. We have thus found that, in comparison with the case of the full system of equations,
 the solutions of the DNLS equation are unique
 for each choice of the coupling constant.

\subsubsection{Comparison of results}
Since we have found that in the stationary limit the full system of equations
can be replaced by a DNLS equation, the minimal energy 
solutions we have obtained for both types of equations should be in agreement. 

Comparing the two systems, we see that the value $\tilde{g}$ is given in terms of the coupling constants
of the full system by:
\begin{equation}
\tilde{g}=\frac{2}{9}\frac{c_x^2}{M j_x k_x}
\end{equation}
which, for the choice of coupling constants we have used in our numerical simulations,
gives:
\begin{equation}
\tilde{g}=\frac{c_x^2}{90} \ .
\end{equation}
Thus a critical value of $c_x\approx 20$ would imply $\tilde{g}_{cr}\approx 4.4$.
First, we remark that the values of the critical electron-phonon coupling  
we obtained from all our simulations (including those for
the continuous MNLS equation) are of the same order of magnitude.
However, there is a slight discrepancy between the results for the full system
and the DNLS equation. We believe that this is due to the fact that there might exist
additional terms ${\cal A}$ in (\ref{solution}) for which $\Delta(1) {\cal A}=0$ 
and/or $\Delta(2) {\cal A}=0$. These terms would then appear in (\ref{NLS}) and would
change the comparison of the solutions. However, it is
difficult to determine these additional terms 
 and so this  is left as a future work \cite{hz2}. 

\subsubsection{Tubes with different diameter, chirality and lengths}
Since most of our results are for a $(5,5)$ armchair tube
and since it is well known that the properties of nanotubes depend
strongly on the diameter, chirality and lengths of the tube, we will discuss
briefly how different tubes could be constructed in our model.
We haven't constructed these tubes yet, but we aim to do so in a future publication
in which we intend to extend our approach to a more realistic 
three dimensional model \cite{hz2}.

Labelling the first carbon atom in $y$ direction by $j=0$, we have chosen
$j_{max}$ such that it is divisable by $4$. Thus, the length of the
tube in the $y$-direction is $l_y=\frac{3}{4}j_{max}$. Since we identify the
fields labelled by $(i=0,j)$ with those at $(i=i_{max},j)$, the diameter
of the tube is given by $d=l_y/\pi$. Thus increasing/decreasing $j_{max}$
by $4n$, $n=1,2,3...$, we can construct armchair nanotubes
with diameters $d=\frac{3}{4\pi} (j_{max}\pm 4n)$. Similarly, we can construct
longer tubes by increasing the number of atoms in the $x$ direction.

As far as chirality is concerned, there are two things to modify in our model
in order to be able to 
construct tubes with different chirality. One is to change the number
of points in the $y$ direction so that $j_{max}$ is non-divisable by $4$.
The other is to adjust the periodic boundary conditions in the $y$-direction
appropriately.
If we e.g. choose $j_{max}=18$, we have to identify the fields
at $(i=0,j)$ with those at $(i=i_{max},j+1)$. This then would give us
a $(5,4)$ nanotube which would be semi-conducting.

In this work, we have concentrated our attention on the existence of localised structures.
These structures extend over large parts of our grids, but are
negligible at boundaries. Hence, we expect these to hold for systems with different
boundary conditions, i.e. different chiralities.

\section{Conclusions}
Motivated by a large amount of research done in the area of nanotubes, we have studied
solitons on a $2$-dimensional hexagonal lattice. We have chosen our lattice to
be periodic in both the $x$ and $y$ directions and to be of large extention in one (the $x$) direction
and of small extention in the other (the $y$) direction. In the stationary limit, 
we have found that the full system of equations in which the electron excitation  is coupled to
the displacement fields of the lattice can be replaced by a modified discrete non-linear
Schr\"odinger (DNLS) equation. This discovery of an {\bf exact} solution of the
full system of equations is remarkable since for the similar quadratic lattice such a simple solution does not exist.

In our numerical studies we  have mainly concentrated our attention on determining the value of the
critical phonon-electron coupling constant. For the DNLS we have found that unique solutions
exist and that the value of the critical coupling is in good agreement with both the 
analytically and numerically values found for the continuous analog of the DNLS. For the full system of
equations, we believe that a large number of solutions exist for each choice of
the coupling constants and that a conserved quantity exists in the system.
The critical value of the electron-phonon coupling is of the same order of
magnitude as in the case of the DNLS; however, we believe that this small
discrepancy results from the fact that possible ``boundary'' terms appear
when replacing the full system by the DNLS. These boundary terms are
terms which are annihilated by either the $4$-point Laplacian $\Delta(1)$ and/or
by the $7$-point Laplacian $\Delta(2)$. To find these terms is non-trivial
and since this seems an interesting topic by itself, we leave this as a future work \cite{hz2}.

Finally let us mention that a possible extension of the results given here
would involve the study of the corresponding three dimensional equations
and/or of the influence of external forces.\\ 
\\
\\
{\bf Acknowledgements} BH was supported by an EPSRC grant.

\newpage

\section{Appendix}
\subsection{Hamiltonian and Equations of motion for the numerical studies}
To simplify the numerical construction of the solutions we have 
squeezed the lattice as indicated in Fig.s~2a, 2b. This reduces the memory
requirements and so speeds up the calculations.
The Hamiltonian $H^{n}$ for the numerical construction thus takes the form

\begin{eqnarray}
H^{n} &=&
\sum_{j=1}^{N_2} \sum_{\frac{i-1}{4}=0}^{\frac{N_1}{4}-3}\left[ (E+W)\psi_{i,j}\psi^*_{i,j}-
j_x\psi^*_{i,j}\left(\psi_{i+1,j}+\psi_{i-1,j}+\psi_{i+1,j-1}\right) \right. \nonumber \\
&-&
\left. j_x\psi_{i,j}\left(\psi^*_{i+1,j}+\psi^*_{i-1,j}+\psi^*_{i+1,j-1}\right) \right. \nonumber \\
&+&
\left. \psi_{i,j}\psi^*_{i,j}\left(\frac{c_x}{3}(u_{i+1,j}+u_{i+1,j-1}-2u_{i-1,j})
-\frac{c_x}{\sqrt{3}}(v_{i+1,j-1}-v_{i+1,j})\right) \right] \nonumber \\
&+&
\sum_{j=1}^{N_2} \sum_{\frac{i-2}{4}=0}^{\frac{N_1}{4}-2} \left[ (E+W)\psi_{i,j}\psi^*_{i,j}-
j_x\psi^*_{i,j}\left(\psi_{i+1,j}+\psi_{i-1,j}+\psi_{i-1,j+1}\right)\right. \nonumber \\
&-&
\left. j_x\psi_{i,j}\left(\psi^*_{i+1,j}+\psi^*_{i-1,j}+\psi^*_{i-1,j+1}\right) \right.\nonumber \\
&+&
\left. \psi_{i,j}\psi^*_{i,j}\left(\frac{c_x}{3}(-u_{i-1,j}-u_{i-1,j+1}+2u_{i+1,j})
+\frac{c_x}{\sqrt{3}}(v_{i-1,j+1}-v_{i-1,j})\right) \right] \nonumber \\
&+&
\sum_{j=1}^{N_2} \sum_{\frac{i-3}{4}=0}^{\frac{N_1}{4}-1} \left[ (E+W)\psi_{i,j}\psi^*_{i,j}-
j_x\psi^*_{i,j}\left(\psi_{i+1,j}+\psi_{i-1,j}+\psi_{i+1,j+1}\right) \right. \nonumber \\
&-&
\left. 
j_x\psi_{i,j}\left(\psi^*_{i+1,j}+\psi^*_{i-1,j}+\psi^*_{i+1,j+1}\right) \right. \nonumber \\
&+&
\left. 
\psi_{i,j}\psi^*_{i,j}\left(\frac{c_x}{3}(u_{i+1,j}+u_{i+1,j+1}-2u_{i-1,j})
+\frac{c_x}{\sqrt{3}}(v_{i+1,j+1}-v_{i+1,j})\right) \right]\nonumber \\
&+&
\sum_{j=1}^{N_2} \sum_{\frac{i}{4}=1}^{\frac{N_1}{4}} \left[ (E+W)\psi_{i,j}\psi^*_{i,j}-
j_x\psi^*_{i,j}\left(\psi_{i+1,j}+\psi_{i-1,j}+\psi_{i-1,j-1}\right) \right. \nonumber \\
&-&
\left.
j_x\psi_{i,j}\left(\psi^*_{i+1,j}+\psi^*_{i-1,j}+\psi^*_{i-1,j-1}\right) \right. \nonumber \\
&+&
\left. 
\psi_{i,j}\psi^*_{i,j}\left(\frac{c_x}{3}(u_{i-1,j}+u_{i-1,j-1}-2u_{i+1,j})
+\frac{c_x}{\sqrt{3}}(v_{i-1,j}-v_{i-1,j-1})\right) \right]
\end{eqnarray}
with the phonon energy $W^n$~:
\begin{eqnarray}
W^n&=&\frac{1}{2}M \sum_{j=1}^{N_2} \sum_{i=1}^{N_1} \left( (\frac{du}{dt})^2
+(\frac{dv}{dt})^2+
k_x[(u_{ij}-u_{i-1,j})^2+(v_{ij}-v_{i-1,j})^2] \right)\nonumber \\
&+&
\frac{1}{2}M\sum_{j=1}^{N_2}\sum_{\frac{i-2}{4}=0}^{\frac{N_1}{4}-2}  \left(
k_x[(u_{ij}-u_{i-1,j+1})^2+(v_{ij}-v_{i-1,j+1})^2] \right)\nonumber \\
&+&
\frac{1}{2}M\sum_{j=1}^{N_2} \sum_{\frac{i-3}{4}=0}^{\frac{N_1}{4}-1} \left(
k_x[(u_{ij}-u_{i+1,j+1})^2+(v_{ij}-v_{i+1,j+1})^2] \right) \ .
\end{eqnarray}
The equations of motion are then given by
\begin{itemize}
\item For $i=1+4k$, $k=1,2...$:
\begin{eqnarray}
i\hbar \frac{\partial \psi_{i,j}}{\partial t}&=&(E+W)\psi_{i,j}-2j_{x}
\left(\psi_{i+1,j}+\psi_{i-1,j}+\psi_{i+1,j-1}\right) \nonumber \\
&+&\psi_{i,j}\left[\frac{c_{x}}{3}(u_{i+1,j}+u_{i+1,j-1}-2 u_{i-1,j})
+\frac{c_{x}}{\sqrt{3}}(v_{i+1,j}-v_{i+1,j-1})
\right]
\end{eqnarray}
\begin{eqnarray}
\frac{d^2 u_{i,j}}{dt^2}&=&-k_{x}\left(3u_{i,j}-u_{i+1,j}-u_{i-1,j}-u_{i+1,j-1}\right)\nonumber \\
&-& \frac{c_{x}}{3M}\left(2\psi_{i-1,j} \psi^*_{i-1,j}-\psi_{i+1,j} \psi^*_{i+1,j}
-\psi_{i+1,j-1} \psi^*_{i+1,j-1}\right) 
\end{eqnarray}
\begin{eqnarray}
\frac{d^2 v_{i,j}}{dt^2}&=&-k_{x}\left(3v_{i,j}-v_{i+1,j}-v_{i-1,j}-v_{i+1,j-1}\right)\nonumber \\
&-& \frac{c_{x}}{\sqrt{3}M}\left(\psi_{i+1,j-1} 
\psi^*_{i+1,j-1}-\psi_{i+1,j} \psi^*_{i+1,j}\right) 
\end{eqnarray}

\item For $i=2+4k$, $k=1,2...$:
\begin{eqnarray}
i \hbar\frac{\partial \psi_{i,j}}{\partial t}&=&(E+W)\psi_{i,j}-2j_{x}
\left(\psi_{i+1,j}+\psi_{i-1,j}+\psi_{i-1,j+1}\right) \nonumber \\
&+&\psi_{i,j}\left[\frac{c_{x}}{3}(-u_{i-1,j}-u_{i-1,j+1}+2 u_{i+1,j})
+\frac{c_{x}}{\sqrt{3}}(v_{i-1,j+1}-v_{i-1,j})
\right]
\end{eqnarray}
\begin{eqnarray}
\frac{d^2 u_{i,j}}{dt^2}&=&-k_{x}\left(3u_{i,j}-u_{i+1,j}-u_{i-1,j}-u_{i-1,j+1}\right)\nonumber \\
&+& \frac{c_{x}}{3M}\left(2\psi_{i+1,j} \psi^*_{i+1,j}-\psi_{i-1,j} \psi^*_{i-1,j}
-\psi_{i-1,j+1} \psi^*_{i-1,j+1}\right) 
\end{eqnarray}
\begin{eqnarray}
\frac{d^2 v_{i,j}}{dt^2}&=&-k_{x}\left(3v_{i,j}-v_{i+1,j}-v_{i-1,j}-v_{i-1,j+1}\right)\nonumber \\
&+& \frac{c_{x}}{\sqrt{3}M}\left(\psi_{i-1,j+1} 
\psi^*_{i-1,j+1}-\psi_{i-1,j} \psi^*_{i-1,j}\right) 
\end{eqnarray}

\item For $i=3+4k$, $k=1,2...$:
\begin{eqnarray}
i \hbar\frac{\partial \psi_{i,j}}{\partial t}&=&(E+W)\psi_{i,j}-2j_{x}
\left(\psi_{i+1,j}+\psi_{i-1,j}+\psi_{i+1,j+1}\right) \nonumber \\
&+&\psi_{i,j}\left[\frac{c_{x}}{3}(u_{i+1,j}+u_{i+1,j+1}-2 u_{i-1,j})
+\frac{c_{x}}{\sqrt{3}}(v_{i+1,j+1}-v_{i+1,j})
\right]
\end{eqnarray}
\begin{eqnarray}
\frac{d^2 u_{i,j}}{dt^2}&=&-k_{x}\left(3u_{i,j}-u_{i+1,j}-u_{i-1,j}-u_{i+1,j+1}\right)\nonumber \\
&-& \frac{c_{x}}{3M}\left(2\psi_{i-1,j} \psi^*_{i-1,j}-\psi_{i+1,j} \psi^*_{i+1,j}
-\psi_{i+1,j+1} \psi^*_{i+1,j+1}\right) 
\end{eqnarray}
\begin{eqnarray}
\frac{d^2 v_{i,j}}{dt^2}&=&-k_{x}\left(3v_{i,j}-v_{i+1,j}-v_{i-1,j}-v_{i+1,j+1}\right)\nonumber \\
&-& \frac{c_{x}}{\sqrt{3}M}\left(\psi_{i+1,j} 
\psi^*_{i+1,j}-\psi_{i+1,j+1} \psi^*_{i+1,j+1}\right) 
\end{eqnarray}

\item For $i=4+4k$, $k=1,2...$:
\begin{eqnarray}
i \hbar\frac{\partial \psi_{i,j}}{\partial t}&=&(E+W)\psi_{i,j}-2j_{x}
\left(\psi_{i+1,j}+\psi_{i-1,j}+\psi_{i-1,j-1}\right) \nonumber \\
&-&\psi_{i,j}\left[\frac{c_{x}}{3}(u_{i-1,j}+u_{i-1,j-1}-2 u_{i+1,j})
-\frac{c_{x}}{\sqrt{3}}(v_{i-1,j}-v_{i-1,j-1})
\right]
\end{eqnarray}
\begin{eqnarray}
\frac{d^2 u_{i,j}}{dt^2}&=&-k_{x}\left(3u_{i,j}-u_{i+1,j}-u_{i-1,j}-u_{i-1,j-1}\right)\nonumber \\
&+& \frac{c_{x}}{3M}\left(2\psi_{i+1,j} \psi^*_{i+1,j}-\psi_{i-1,j} \psi^*_{i-1,j}
-\psi_{i-1,j-1} \psi^*_{i-1,j-1}\right) 
\end{eqnarray}
\begin{eqnarray}
\frac{d^2 v_{i,j}}{dt^2}&=&-k_{x}\left(3v_{i,j}-v_{i+1,j}-v_{i-1,j}-v_{i-1,j-1}\right)\nonumber \\
&+& \frac{c_{x}}{\sqrt{3}M}\left(\psi_{i-1,j} 
\psi^*_{i-1,j}-\psi_{i-1,j-1} \psi^*_{i-1,j-1}\right) 
\end{eqnarray}
\end{itemize}

\newpage
\begin{fixy}{-1}
\begin{figure}
\centering
\epsfysize=10cm
\mbox{\epsffile{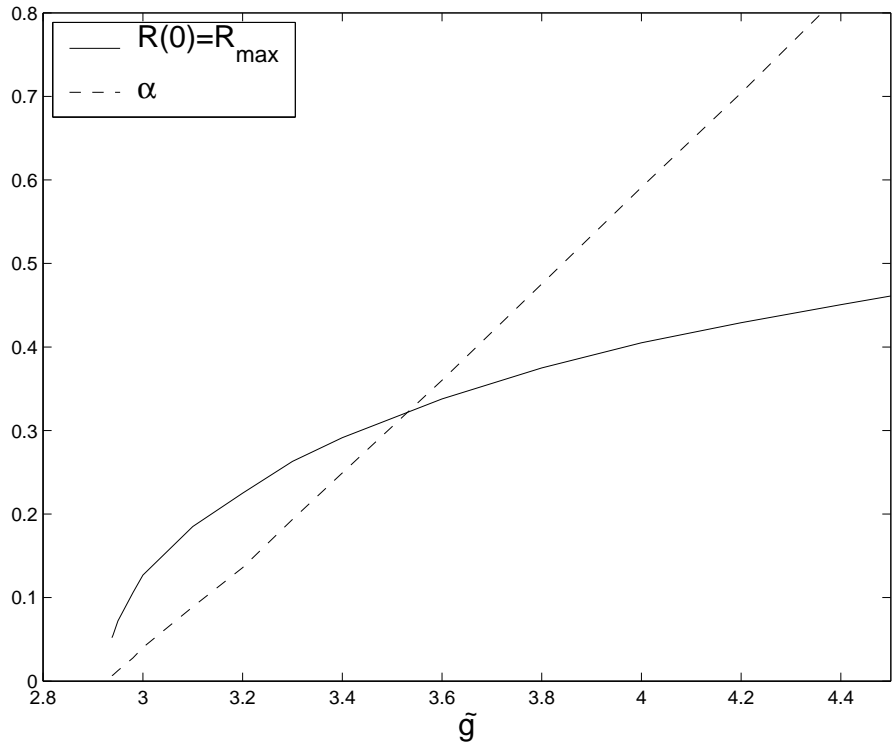}}
\label{fig1}
\caption{\label{Fig.1(a)} The results obtained with the ansatz $\psi(r,\theta)=e^{i\alpha t} R(r)$ to solve
the continuous modified NLS equation. In this figure, the value of the function $R(r)$ at the origin, $R(0)$, which
in our construction is equal to the maximum of $R$, $R_{max}$, is shown as
function of $\tilde{g}$. The values of $\alpha$ are also shown. The solutions shown have norm  one.}
\end{figure}
\end{fixy}

\newpage
\begin{fixy}{0}
\begin{figure}
\centering
\epsfysize=9cm
\mbox{\epsffile{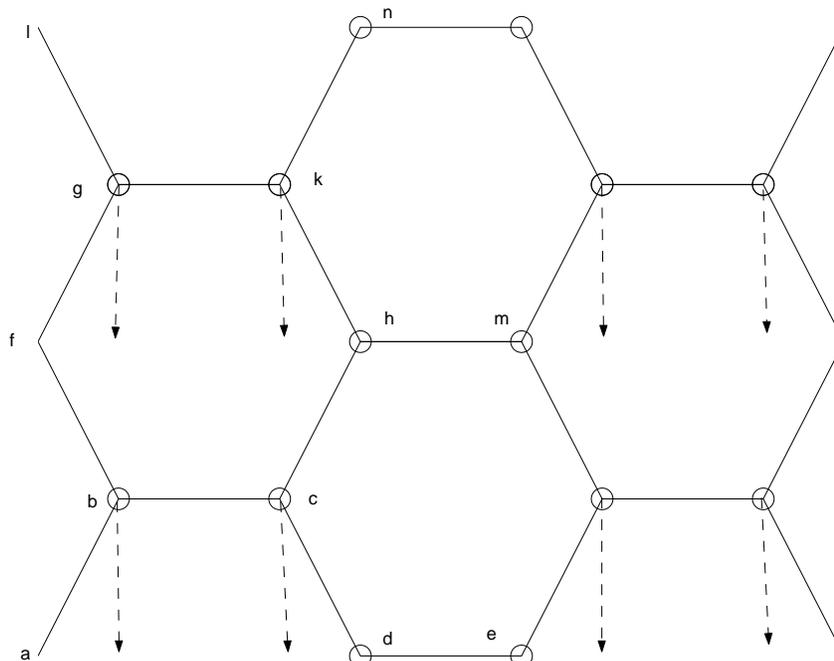}}
\label{fig2}
\caption{\label{Fig.2(a)} The hexagonal lattice is shown. The arrows indicate the method
of ``squeezing'' the lattice for the numerical evaluation.   }
\end{figure}

\begin{figure}
\centering
\epsfysize=9cm
\mbox{\epsffile{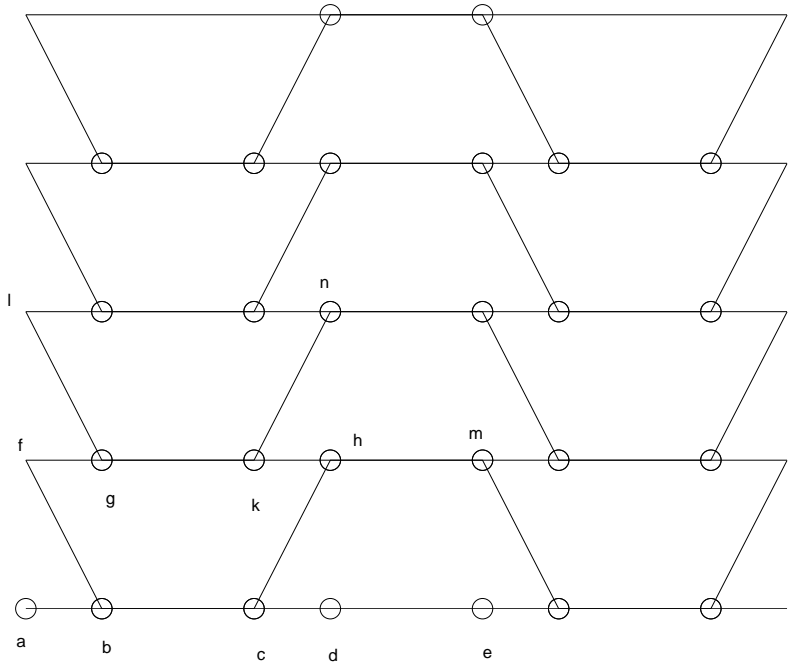}}
\caption{\label{Fig.2(b)} The ``squeezed'' hexagonal lattice for the numerical
construction. The Hamiltonian $H^n$ corresponding 
to this lattice is given in the Appendix of the paper. }
\end{figure}
\end{fixy}
\newpage

\newpage
\begin{fixy}{-1}
\begin{figure}
\centering
\epsfysize=10cm
\mbox{\epsffile{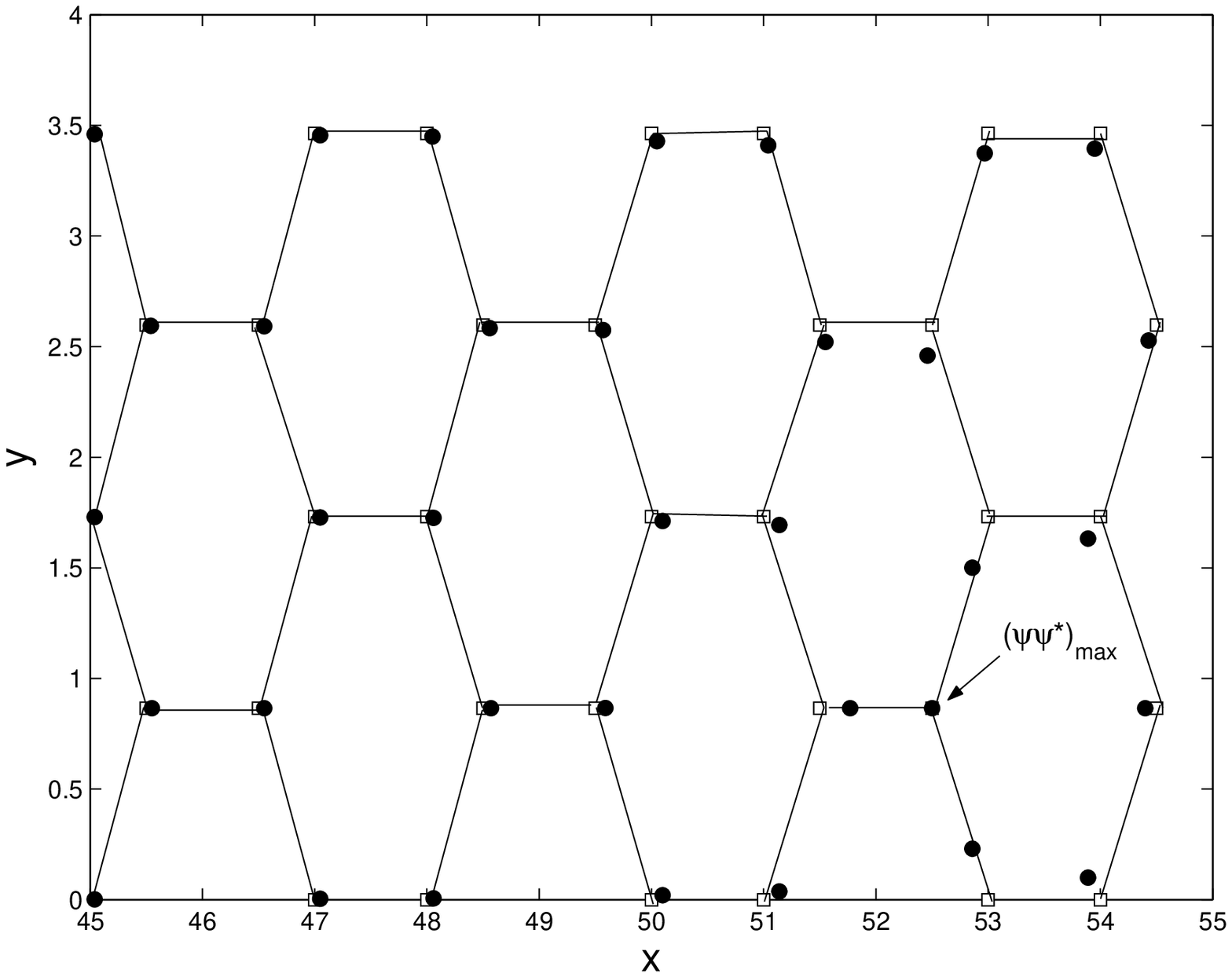}}
\label{fig3}
\caption{\label{Fig.3} The distortion of the lattice close to the location of the
soliton is shown.
The squares indicate the undistorted lattice, while the circles indicate the distorted lattice
after $t=4000$; $c_x=25$. The corresponding soliton's maximum $(\psi\psi^*)_{max}\approx 0.6145$.}
\end{figure}
\end{fixy}

\newpage
\begin{fixy}{-1}
\begin{figure}
\centering
\epsfysize=10cm
\label{fig4}
\mbox{\epsffile{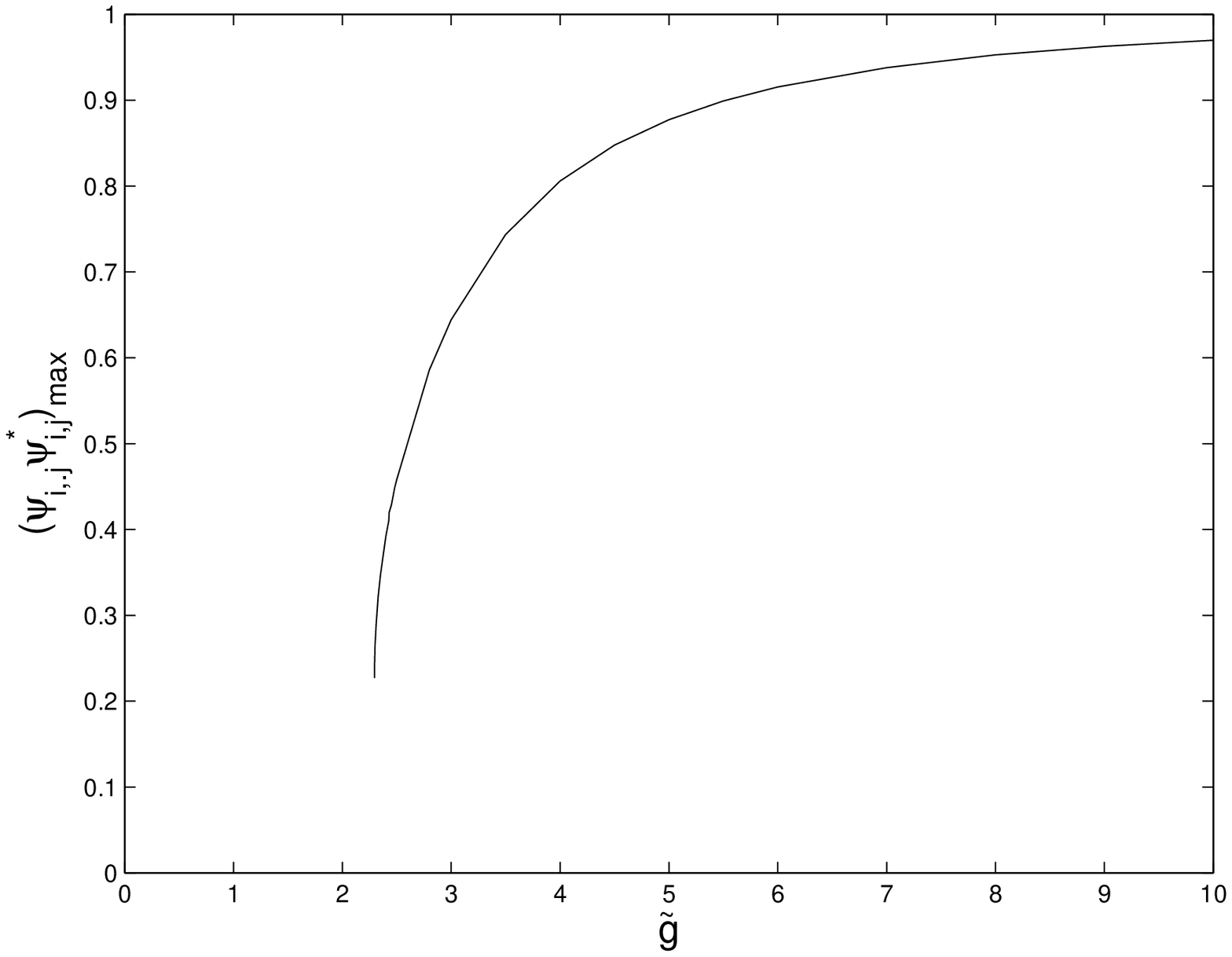}}
\caption{\label{Fig.4} The height of the soliton's maximum $(\psi_{i,j}\psi_{i,j}^*)_{max}$ 
is shown as a function of the
parameter $\tilde{g}$.}
\end{figure}
\end{fixy}


\newpage
\begin{thebibliography}{ccc}
\bibitem{iijima} S. Iijima, Nature {\bf 354} (1991), 56.
\bibitem{nano} see e.g.  M. S. Dresselhaus, G. Dresselhaus and P. Eklund,
{\it The science of fullerenes and carbon nanotubes}, Academic (1996);
 {\it Carbon nanotubes, preparation and properties}, 
edited by T.W.Ebbesen, CRC Press (1996);
R. Saito, G. Dresselhaus and M. S. Dresselhaus, {\it Physical properties
of carbon nanotubes}, World Scientific (1998);
P. J. F. Harris, {\it Carbon nanotubes and related
structures}, Cambridge University Press (1999);
{\it Carbon Nanotubes: Synthesis, Structure, Properties, and Applications},
edited by M. S. Dresselhaus, G. Dresselhaus and P. Avouris, Springer-Verlag
(2000).
\bibitem{cn} M. S. C. Mazzoni and H. Chacham, Phys. Rev. {\bf B61} (2000), 7312;
L. Yang, M. P. Anantram, J. Han and J. P. Lu, Phys. Rev. {\bf B60} (1999), 13874;
C.-J. Park, Y.-H. Kim, K. J. Chang, Phys. Rev. {\bf B60} (1999), 10656;
M. S. C. Mazzoni and H. Chacham, Appl. Phys. Lett. {\bf 76} (2000), 1561;
M. Verissimo-Alves, R. B. Capaz, B. Koiller, E. Artacho and H. Chacham, Phys. Rev. Lett. {\bf 86} (2001), 3372.
\bibitem{davy} A. S. Davydov, {\it Solitons in molecular systems}, Reidel,
Dordrecht (1985).
\bibitem{review} A. Scott, Phys. Rep. {\bf 217} (1992), 1;
{\it Nonlinear excitations in Biomolecules}, Ed.: M. Peyrard, Springer, Berlin (1996).
\bibitem{bpz1} L. Brizhik, A. Eremko, B. Piette and W. J. Zakrzewski,
Physica {\bf D 146} (2000), 275.
\bibitem{bpz3} L. Brizhik, B. Piette and W. J. Zakrzewski, Ukr. Fiz. Journal {\bf 46} (2001), 503.
\bibitem{bpz2} L. Brizhik, A. Eremko, B. Piette and W. J. Zakrzewski,
Physica {\bf D 159} (2001), 71.
\bibitem{liu} L. Liu, G. Y. Guo, C. S. Jayanthi and S. Y. Wu, Phys. Rev. Lett. {\bf 88} (2002), 217206.
\bibitem{col} U. Asher, J. Christiansen and R. D. Russell, {\it A collocation solver for fixed
order systems of boundary value problems}, Mathematics of Computation, {\bf 33} (1979), 659;
 U. Asher, J. Christiansen and R. D. Russell, {\it Collocation software for boundary-value ODEs},
ACM Transactions {\bf 7} (1981), 209.
\bibitem{qballs} S. Coleman, Nucl. Phys. {\bf B262} (1985), 263; for a review see e.g.
T. D. Lee and Y. Pang, Phys. Rep. {\bf 221} (1992), 251. 
\bibitem{cbon} T. Spires and R. M. Brown,
{\it ``High Resolution TEM Observations of Single-Walled Carbon Nanotubes''},   
Jr. Department of Botany, The University
of Texas at Austin, Austin, Texas, 78713 (1996),
{\it http://www.botany.utexas.edu/facstaff/facpages
/mbrown/ongres/tspires/nano.htm}; 
J. W. G. Wilder, L. C. Venema, A. G. Rinzler, R. E.
Smalley and C. Dekker, Nature {\bf 391} (1998), 59. 
\bibitem{hz2} B. Hartmann and W. J. Zakrzewski, {\it in preparation}.

\end{thebibliography}
\end{document}